\documentclass[twocolumn,showpacs,pre,letter]{revtex4}

\usepackage{graphicx}

\usepackage{amsmath}

\usepackage{amssymb}

\usepackage{natbib}

\begin{document}

\title{Force heterogeneities in particle assemblies: From order to disorder}

\author{Leonardo E.~Silbert}
\email{lsilbert@physics.siu.edu}

\affiliation{Department of Physics, Southern Illinois University,
  Carbondale, IL 62901, U.S.A.}

\begin{abstract}
  
  The effect of increasing structural disorder on the distribution of
  contact forces $P(f)$, inside three dimensional particle assemblies
  is systematically studied using computer simulations of model
  granular packings. Starting from a face-centred cubic array, where
  all contact forces are identical, an increasing number of defects is
  introduced into the assembly, after which the system is then allowed
  to relax into a new mechanically stable state. Three distinct
  protocols for imposing disorder are compared. A quantitative measure
  of the disorder is obtained from distributions of the coordination
  number and three-particle contact angle. The distribution of normal
  contact forces show dramatic qualitative changes with increasing
  disorder. In the regime where the disorder is relatively weak, the
  pressure and the lowest normal mode frequency scale approximately
  linearly in the coordination number, with distance from the
  crystalline state. These results for $P(f)$ are discussed in the
  context of jamming phenomena in glassy and granular materials.

 \end{abstract}
\pacs
{
61.43.-j % Disordered Solids
83.80.Fg % Granular solids
}

\maketitle

Over the past decade or so it has come to light that the way the
contact forces are distributed through a static pile of grains occurs
in a rather inhomogeneous manner compared to that of an equilibrium
liquid, even though the arrangement of the particles are similar in
both cases \cite{bernal5}. Whereas many pairs of touching particles in
a granular packing experience only a small force between them, others
experience much larger forces. In fact, there are more of these
high-force contacts than one might expect if a static packing of
particles were truly representative of a snapshot of a thermodynamic
system. What is becoming increasingly apparent, is that the properties
of granular packings appear to share many similarities with amorphous,
glassy phases of their atomic and colloidal counterparts
\cite{coniglio1}. Simply due to the fact that grains are easier to see
than atoms, granular materials now hold a prominent role in studies of
amorphous systems.

However, what has yet to be determined in detail is how the properties
of a static packing depend on the arrangement of the particles. To
date, most studies have focused on the two extreme cases of either,
fully disordered particle packings, where there are no long-ranged
correlations in the particle positions, or perfectly ordered arrays.
Yet, it is not clear how the properties of a static packing change as
the configuration is tuned from an ordered array to a disordered
state. It is this type of order-disorder ``transition'' that is
addressed here.

With the recent upsurge in the study of granular materials, several
notable properties of particle packings have emerged. One of the most
robust, and reproducible, features of a {\it disordered} grain pile,
that sets it apart from more traditional crystalline solids, is the
probability distribution function of contact forces $P(f)$, where
$f\equiv F/<F>$ is the normal force $F$ between two particles in
contact, normalized by the configuration-averaged force $<F>$. It has
been shown experimentally, numerically, and by computer simulations,
that $P(f)$ for a disordered particle packing exhibits an
exponential'ish \cite{comment2} decay at forces $f$, greater (roughly
twice) than the average force, and a peak or plateau at small forces
(below half the average force)
\cite{coppersmith2,mueth1,lovoll1,radjai3,leo11}. All of these studies
suggest that $P(f)$ is quite insensitive to system size as well
dimensionality. Experiments on three dimensional packings
\cite{nagel5} have reported that $P(f)$ is surprisingly insensitive to
the structure of the packing and particle properties (such as the
friction coefficient). For softer, rubber particles there appears to
be a cross-over to a power-law tail at high forces \cite{nagel4}.
Granular dynamics simulations have shown that for disordered packings,
properties such as friction and inelasticity, have only a subtle
effect on $P(f)$ \cite{leo11}, whereas geometrical features of the
packing, such as coordination number, do depend quite sensitively on
the parameters chosen \cite{leo9}. Hard-sphere simulations of
vacancy-diluted crystals show a depletion of small forces compared
with disordered configurations \cite{torquato5}.

As discussed above, disordered particle packings can exhibit a number
of subtle differences. However, there is one extreme case that is
well-defined. For a perfectly ordered, face-centred cubic (fcc),
crystalline array, all particles have the same number of contact
neighbours so that the coordination number is $z = z_{\rm{fcc}} \equiv
12 $, and experience the same contact forces, $f = 1$. The
distributions of the coordination number $P(z)$, and contact forces
$P(f)$, are thus known exactly,
\begin{equation}
  \begin{array}{ccc}
    P_{\rm{fcc}}(z) & = & \delta(z-z_{\rm{fcc}})\\
    P_{\rm{fcc}}(f) & = & \delta(f-1).
  \end{array}
  \label{eqn1}
\end{equation}
Clearly, there is a need to extract which properties of a static array
of particles are responsible for determining the way forces are
distributed. For this reason, the work presented here provides a
systematic study as to how structural disorder affects the
distribution of forces inside three dimensional ($3D$) particle
assemblies.

The computer experiments reported here are for a model system: three
dimensional, frictionless and non-cohesive, soft-sphere packings, with
periodic boundary conditions. Two particles are defined to be
neighbours and interact through a purely-repulsive, short-range,
potential: $V(r) = k (d-r)^{2}$, when their centre--centre separation
$r < d$, where $d$ is the sum of their radii, and $V(r)=0$, otherwise.
To create the particle configurations with varying amounts of
disorder, a number of protocols were implemented:
\begin{itemize}

\item[(i)]
  
  In the first method, $N=16,384$ particles were arranged into a $3D$,
  fcc array, slightly over-compressed to a packing fraction $\phi =
  0.742$, just above that of a hard sphere fcc array, $\phi_{\rm{fcc}}
  = \sqrt{2}\pi/6$, and with $z = z_{\rm{fcc}}$. Disorder was
  introduced by randomly removing, $0.005 < \delta N < 40 \%$ of the
  particles. To remain at constant packing fraction, the system was
  then ``quenched'' to the same initial $\phi=0.742$, using an energy
  minimization algorithm \cite{recipes}. These are denoted {\it
    quenched packings}. The resulting changes in the final packings
  were then analysed. In the following figures, each data set for a
  particular $\delta N$, was averaged over 5 independent realisations.

\item[(ii)]
  
  The second protocol was similar to the first, but instead of
  removing particles, particles chosen at random had their diameters
  reduced by $10\%$, before being re-quenched. In this case, the
  system was made increasingly bidisperse. In each of the two cases
  (i) and (ii), the same procedure was again repeated using quenched
  molecular dynamics, with $N=16384$ and $N=256,000$ particles.

\item[(iii)]
  
  The final protocol used started with $N=32,000$ particles arranged
  into a fcc array at $\phi_{\rm{fcc}}$, constrained in the vertical
  direction by flat base and a free, top surface, with periodic
  boundary conditions in the horizontal plane \cite{leo9}.  Disorder
  was then introduced by removing particles at random, then allowing
  the assembly to relax under gravity - {\it gravity sedimented
    packings}.

\end{itemize}

The various protocols introduced above create configurations which
are, statistically, very similar. Most of the results presented here
are for the quenched, frictionless packings of protocol $(i)$. To
clarify the following nomenclature, the configurations generated via
protocols (i) and (ii) are labelled $Cn$, where $n$ is an index
indicating the amount of disorder. Larger $n$ correspond to packings
with more defects, either by removing particles as in (i), or changing
the particle size, as in (ii). To provide a comparison, two amorphous
packings (A1 and A2) were generated (at two different packing
fractions as detailed below), using the same interaction potential.

\begin{figure}[!]
\includegraphics[width=8cm]{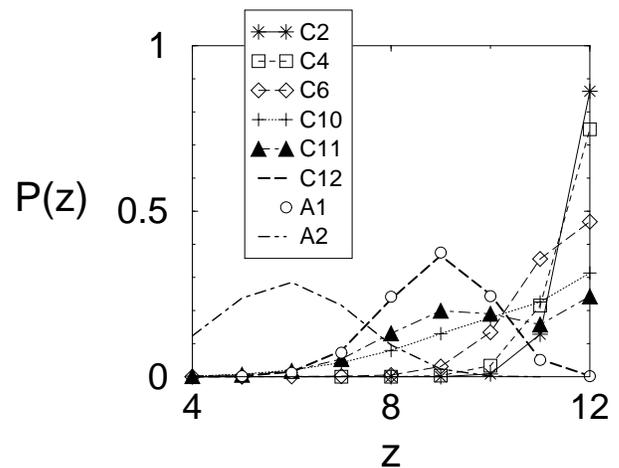}\hfill
\caption{Protocol (i): (a) Coordination number distributions $P(z)$,
  for different amounts of disorder introduced by removing a fraction,
  $\delta N$, of the particles. Increasing configuration label (C)
  corresponds to increasing disorder. The ``A1'' data is an amorphous
  packing generated from an initially random distribution of particles
  quenched to same $\phi=0.742$. For comparison, the ``A2'' data is
  for an amorphous packing at random close packing, $\phi = 0.64$.}
\label{fig1}
\end{figure}

The average coordination number $z$, and the three-particle contact
angle $\theta$, quantify how the protocols generate configurations of
varying disorder. $\theta$ is defined as the angle subtended by
particle $i$ and two of its contact neighbours: $\widehat{jik}$.
Figure \ref{fig1} shows the evolution of the distributions $P(z)$. The
different configurations are labelled in increasing value according to
the number of particles removed: configurations C[1-12] correspond to
$\delta N/N$ = (0.006, 0.012, 0.018, 0.024, 0.030, 0.06, 0.09, 0.12,
0.15, 0.18, 0.24, 0.36). The ``A1'' system was generated from an
initially random distribution of particles quenched to the same
$\phi=0.742$. The ``A2'' amorphous configuration is close to the
random close packing point at $\phi =0.64$. 

\begin{figure}[!]
\includegraphics[width=7cm]{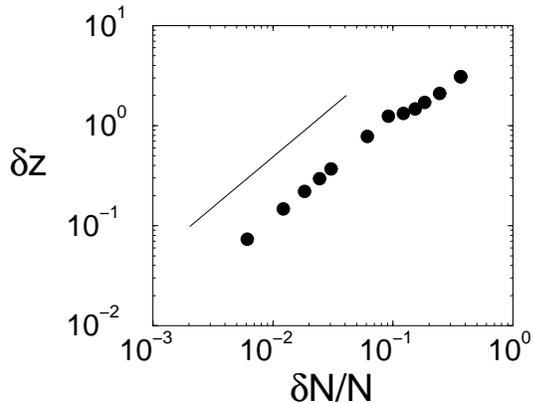}
\caption{Protocol (i): Relation between the number of defects $\delta
  N$, introduced into the packing, and the deviation in the
  coordination number $\delta z$, relative to the initial crystal
  structure. A weakly disordered regime can be identified where,
  $\delta z \propto \delta N$. Here, this corresponds to
  configurations $C[1-7]$.}
\label{fig2}
\end{figure}

Although the $P(z)$ deviate from the fcc delta-function even for
little disorder, the average coordination $z$, remain close to
$z_{\rm{fcc}}$. The deviation in coordination number from the initial
crystal state quantitatively characterises the disorder. As shown in
Fig.~\ref{fig2}, for $\delta N << N$,
\begin{equation}
\delta z \propto \delta N,
\label{eq2}
\end{equation}
where $\delta z \equiv z_{\rm{fcc}} - z$, is the change in the average
coordination number relative to the initial crystal. Thus $\delta z$
provides a convenient measure of how far away the final configuration
is from the crystalline state. Hence, a {\it weakly} disordered regime
can be roughly identified with configuratons for which Eq.~\ref{eq2}
applies, namely configurations $C[1-7]$, for protocol (i).

\begin{figure}[!]
\includegraphics[width=7cm]{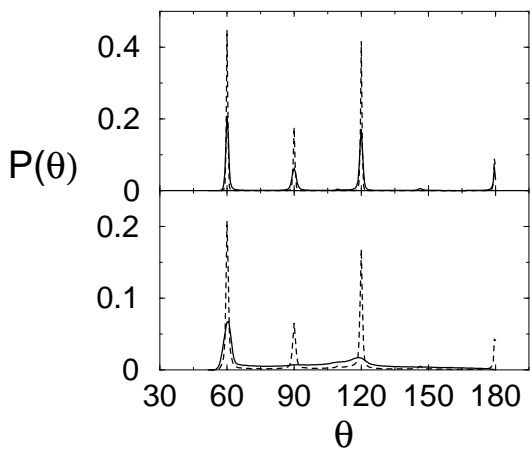}
\caption{Protocol (i): Three-particle contact angle distributions for,
  (top panel) C2 (dashed line) and C6 (full line), and (bottom) C11
  (dashed) and C12 (full).}
\label{fig3}
\end{figure}

Similarly, the distributions in the three-particle contact angle,
$P(\theta)$, shown in Fig.~\ref{fig3}, reflect the fact that protocol
(i) induces only slight deviations from the fcc contact topology. Even
up to C11, there are a significant fraction of three-particle
collineations ($\theta = 180^{\circ}$), whereas, for amorphous
packings these become rare \cite{leo17}.

\begin{figure}[h]
\includegraphics[width=4cm]{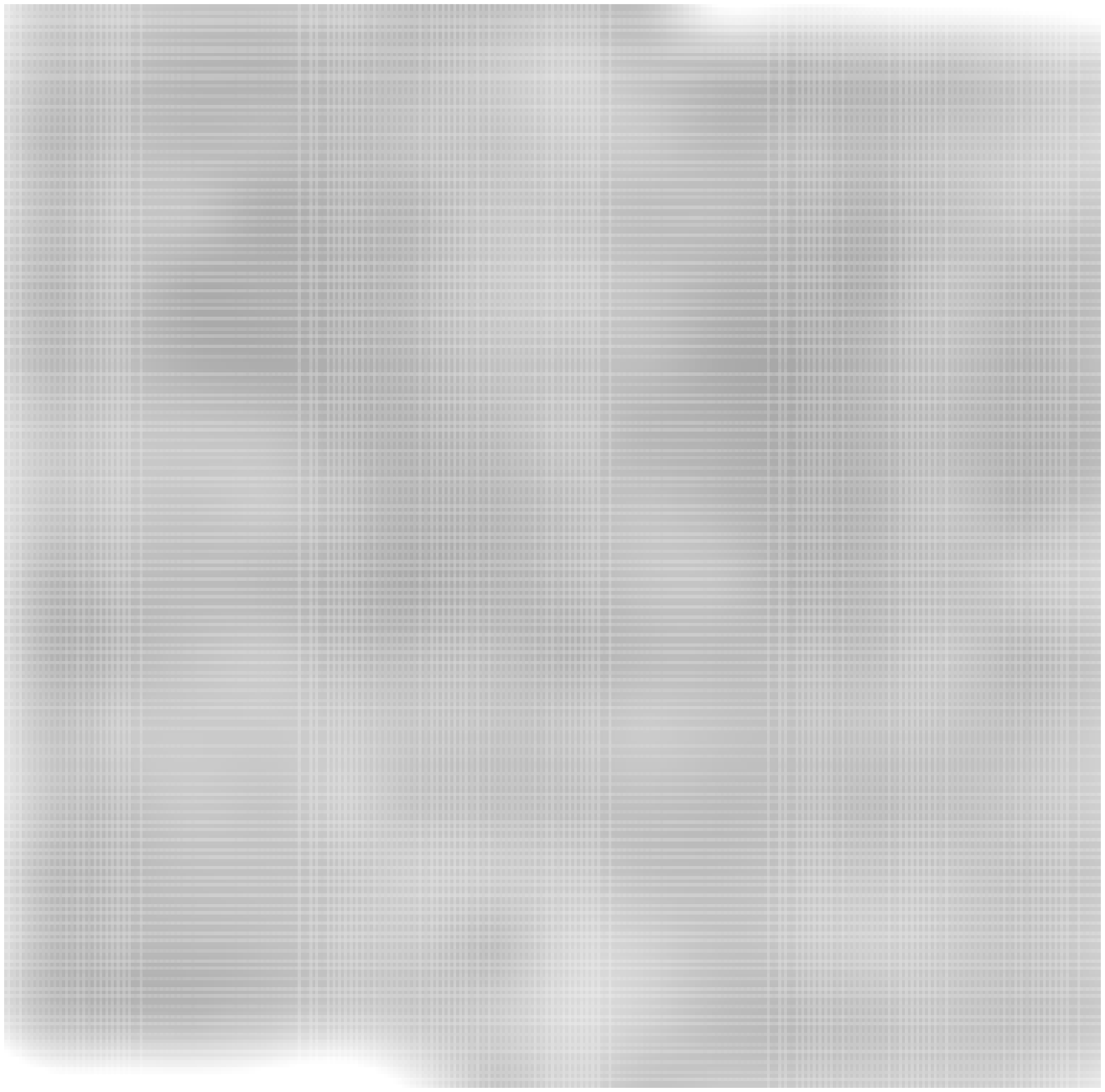}
\includegraphics[width=4cm]{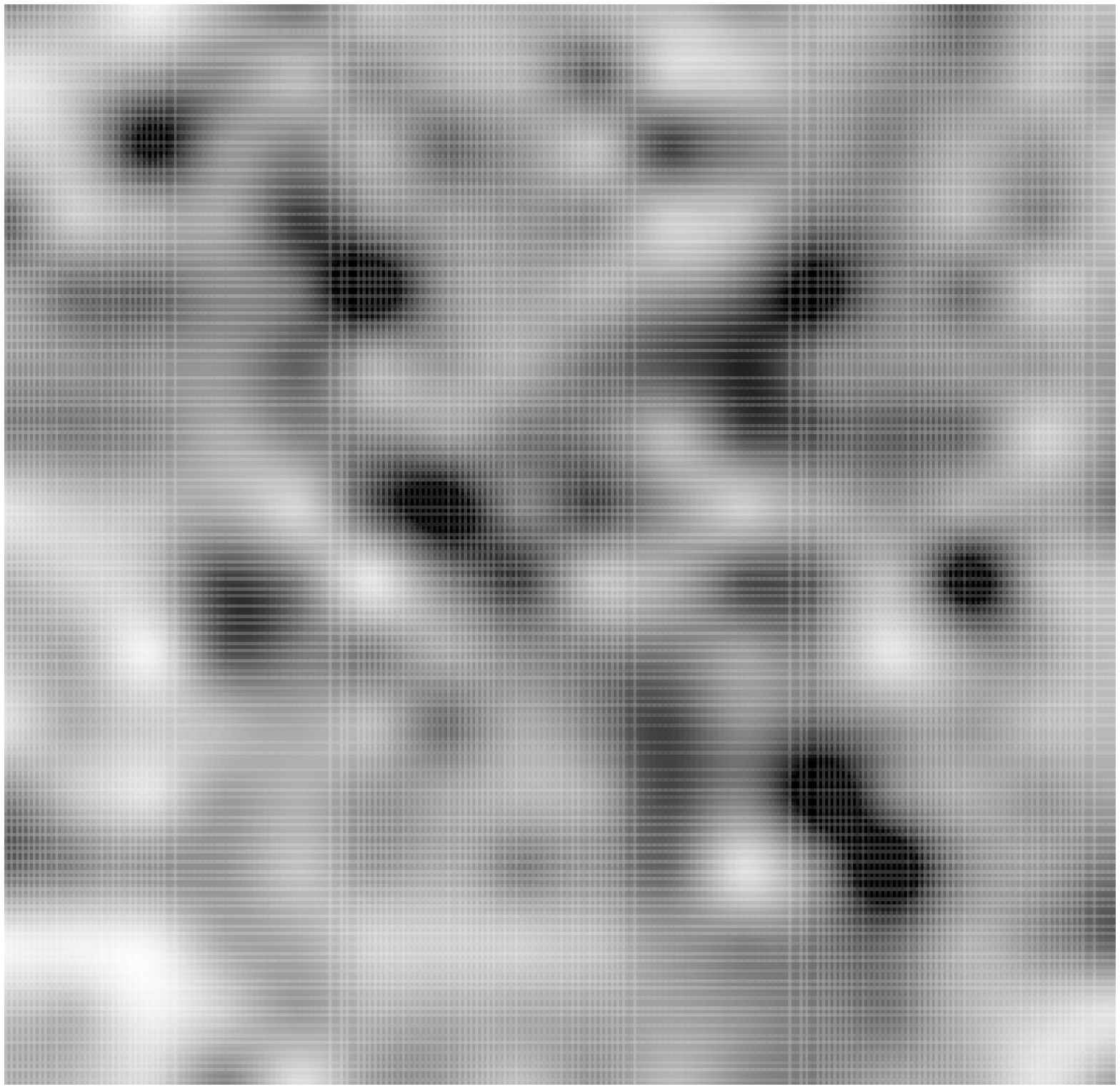}
\caption{Normal stress maps of the relative change in pressure between
  the initial crystal and the disordered packing. Each panel was
  averaged over 5 realisations. Left: Nearly ordered configuration
  (C2). Right: intermediate disorder (C11). Larger stress represented
  by darker shading. Same greyscale in both panels.}
\label{fig4}
\end{figure}

Visualisation of the force networks indicate that the forces are very
sensitive to the imposed disorder. To illustrate this, Fig.~\ref{fig4}
shows the averaged change in the pressure (normal stresses) between
the initial crystal and final the configurations of two systems with
different amounts of disorder.

\begin{figure}[!]
\includegraphics[width=7cm]{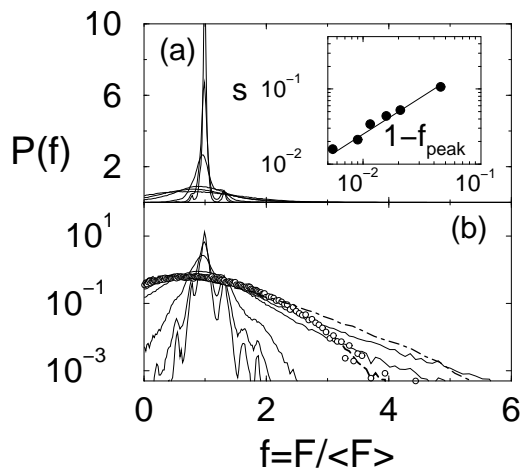}
\caption{Protocol (i): Distributions of normal contact forces $P(f)$,
  for varying disorder at fixed $\phi = 0.742$: (a) linear axes and
  (b) linear-log. In (b), the inner curve is for configuration C2
  (thick solid line), extending outwards with increasing disorder, C4,
  C6, C8, C10, and C12 (thick dashed).  The circles sitting on top of
  C12 is the data for the ``A1'' at $\phi=0.742$. For comparison, the
  thick dash-dotted line is the ``A2'' data at $\phi =0.64$. The inset
  to (a) shows the standard deviations $s$, of Gaussian fits to the
  $P(f)$ peaks, growing linearly with the peak position
  $f_{\rm{peak}}$, shifted relative to the fcc.}
\label{fig5}
\end{figure}

The distribution of normal contact forces $P(f)$ shown in
Fig.~\ref{fig5} quantify the resulting changes in the force network.
For a small fraction of defects, upto and including configuration
$C6$, the region around the primary peaks in $P(f)$ initially undergo
Gaussian broadening from the original, fcc delta-function. Where the
peak remains identifiable, the standard deviations $s$, of Gaussian
fits to the primary $P(f)$ peaks, grow as the location of the peaks
$f_{\rm{peak}}$, move to lower forces with increasing disorder. This
is shown in the inset to Fig.~\ref{fig5}(a). (Recall that for the fcc
array, $f^{\rm{fcc}}_{\rm{peak}} = 1$.) In this regime, the $P(f)$
develop additional features either side of $f_{\rm{peak}}$, reflecting
local force balance in the presence of local defects. These features
correspond to the regime where Eq.~\ref{eq2} apply. As further
disorder is introduced, the distributions become increasingly
asymmetric; showing a dramatic increase in the number of very small
forces and where the high-$f$ tails become increasingly broad
(exponential) \cite{comment3}. Thus, structure plays a crucial role in
determining the heterogeneity of the force network in particle
packings.

\begin{figure}[!]
\includegraphics[width=6cm]{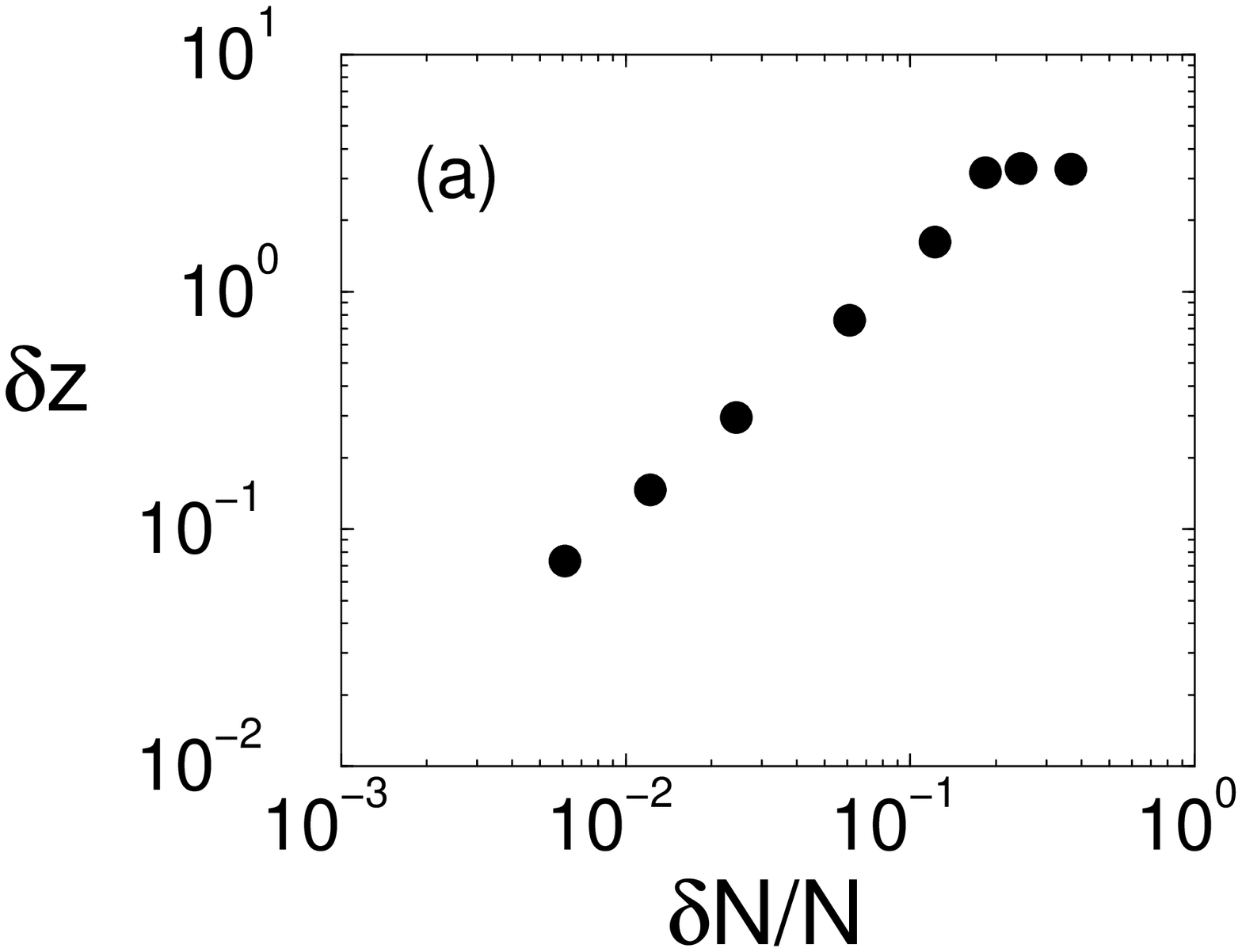}
\includegraphics[width=6cm]{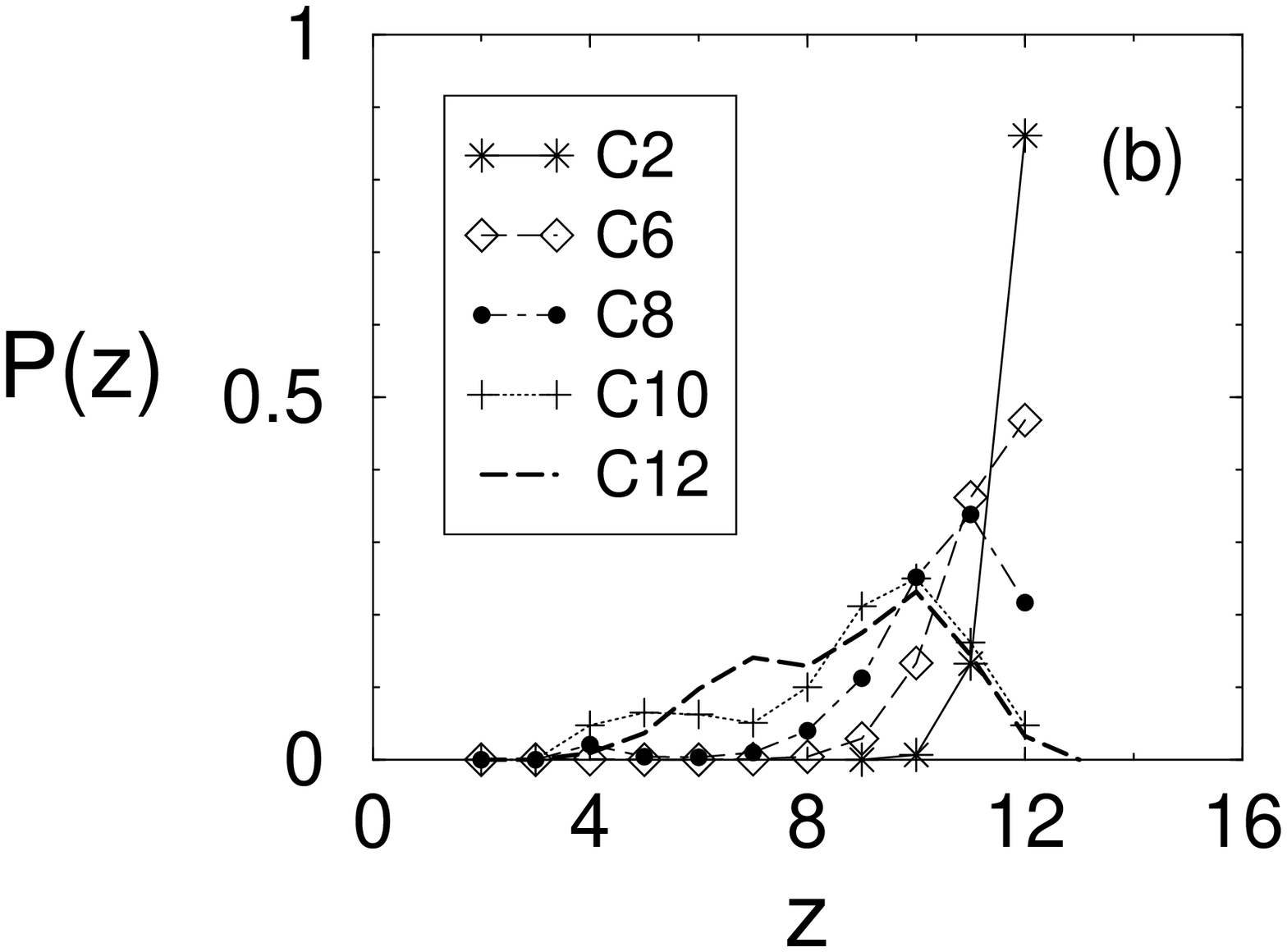}
\includegraphics[width=6cm]{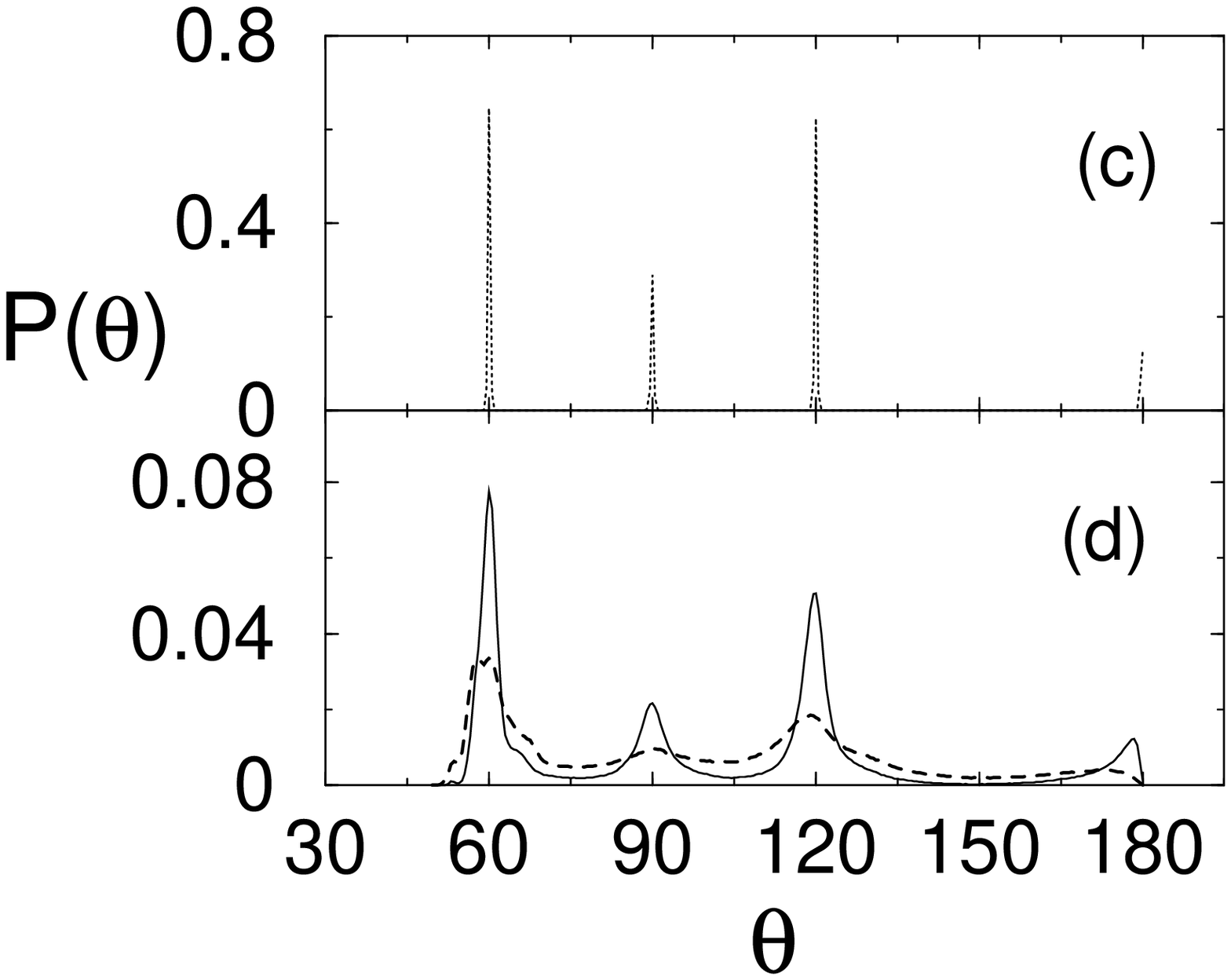}\\
\includegraphics[width=6cm]{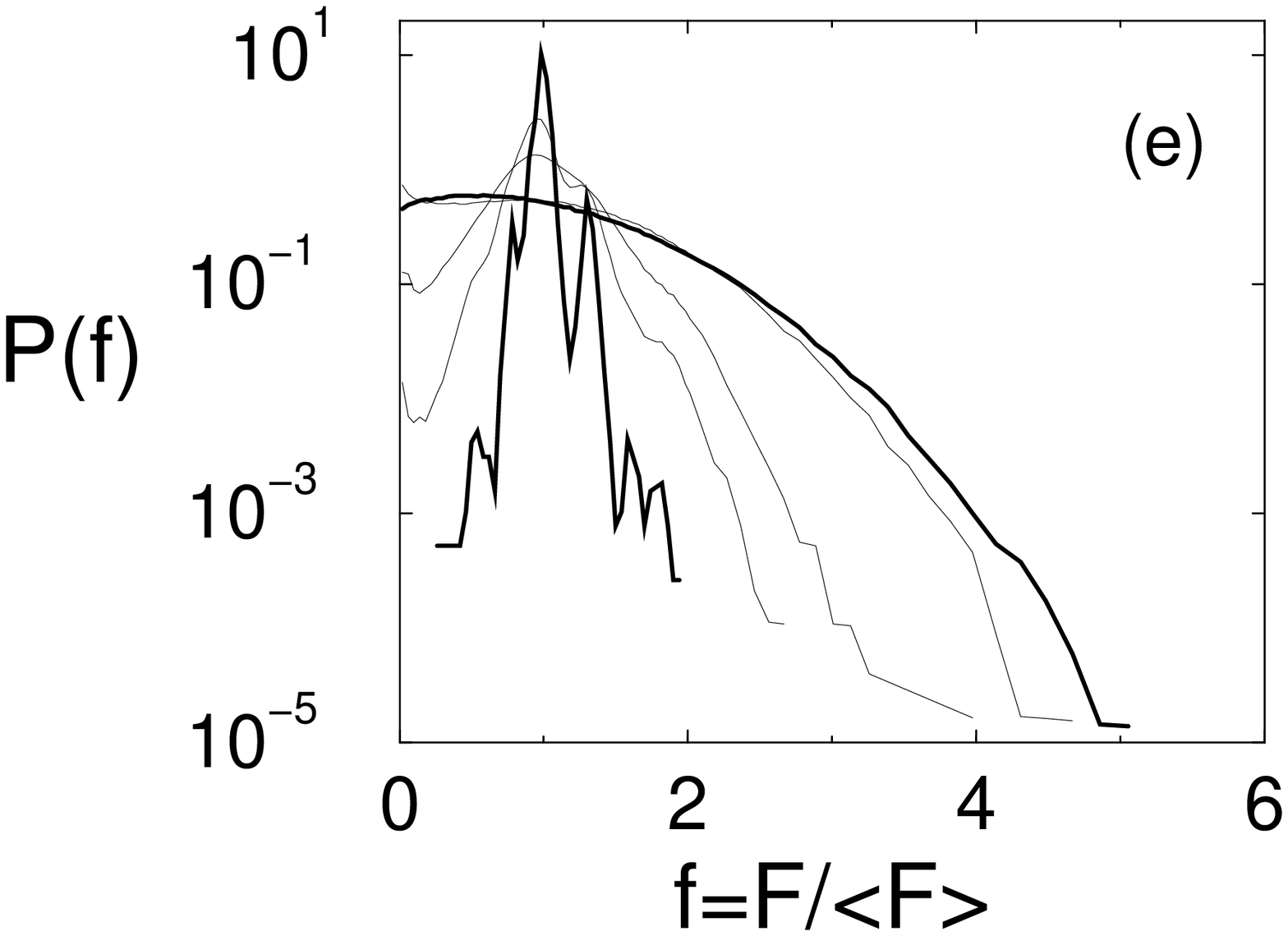}
\caption{Protocol (ii): In, (a) the relative change in the
  coordination number with respect to the initial monodisperse fcc
  array, as the fraction, $\delta N$, of particles are made
  bidisperse. (b) Coordination number distributions $P(z)$, for
  different amounts of disorder at fixed $\phi = 0.742$. Increasing
  configuration label corresponds to increasing disorder. $P(\theta)$
  for, (c) C6, and (d) C10 (solid line) and C12 (dashed). (e) $P(f)$
  over a range of disorder. The inner thick solid line is C2 and the
  outer thick line is C12.}
\label{fig6}
\end{figure}

For small disorder, protocols (i) and (ii) generate configurations
with properties that are almost indistinguishable. For more disorder,
although the configurations generated via protocols (i) and (ii) have
similar properties, there are a few noticeable differences.  For this
reason, the results for protocol (ii) are shown in Fig.~\ref{fig6}.

On comparing the results for protocol (i) (Figs.~\ref{fig1} -
\ref{fig5}), with those of protocol (ii) in Fig.~\ref{fig6}, it
appears that protocol (ii) leads to a more gradual change in the
structure, compared with protocol (i). For (i), there is a much more
dramatic change in structure between C10 and C12, than for the case of
(ii). Still, the general trends remain similar, particularly for
$P(f)$.

\begin{figure}[!]
\includegraphics[width=7cm]{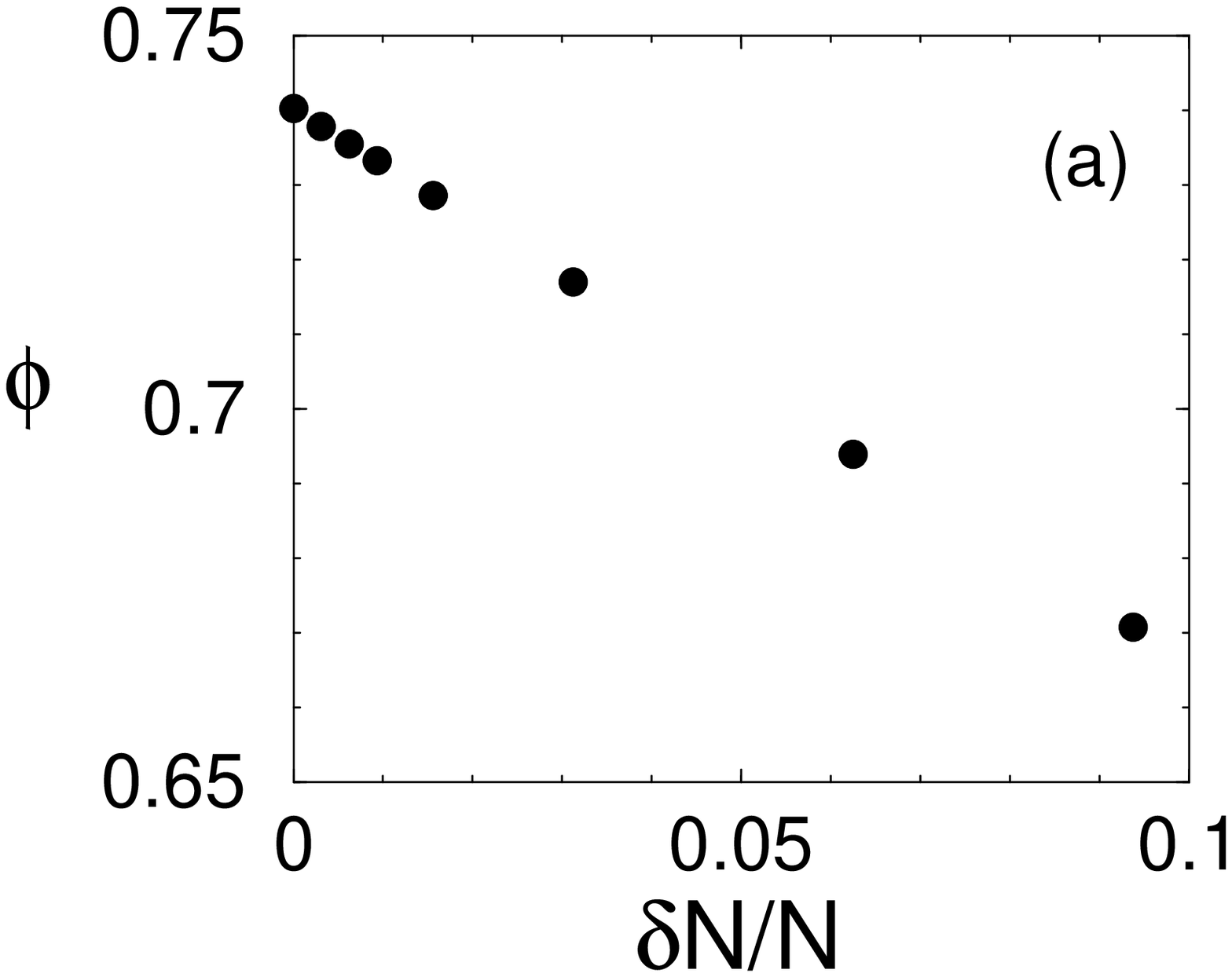}
\includegraphics[width=7cm]{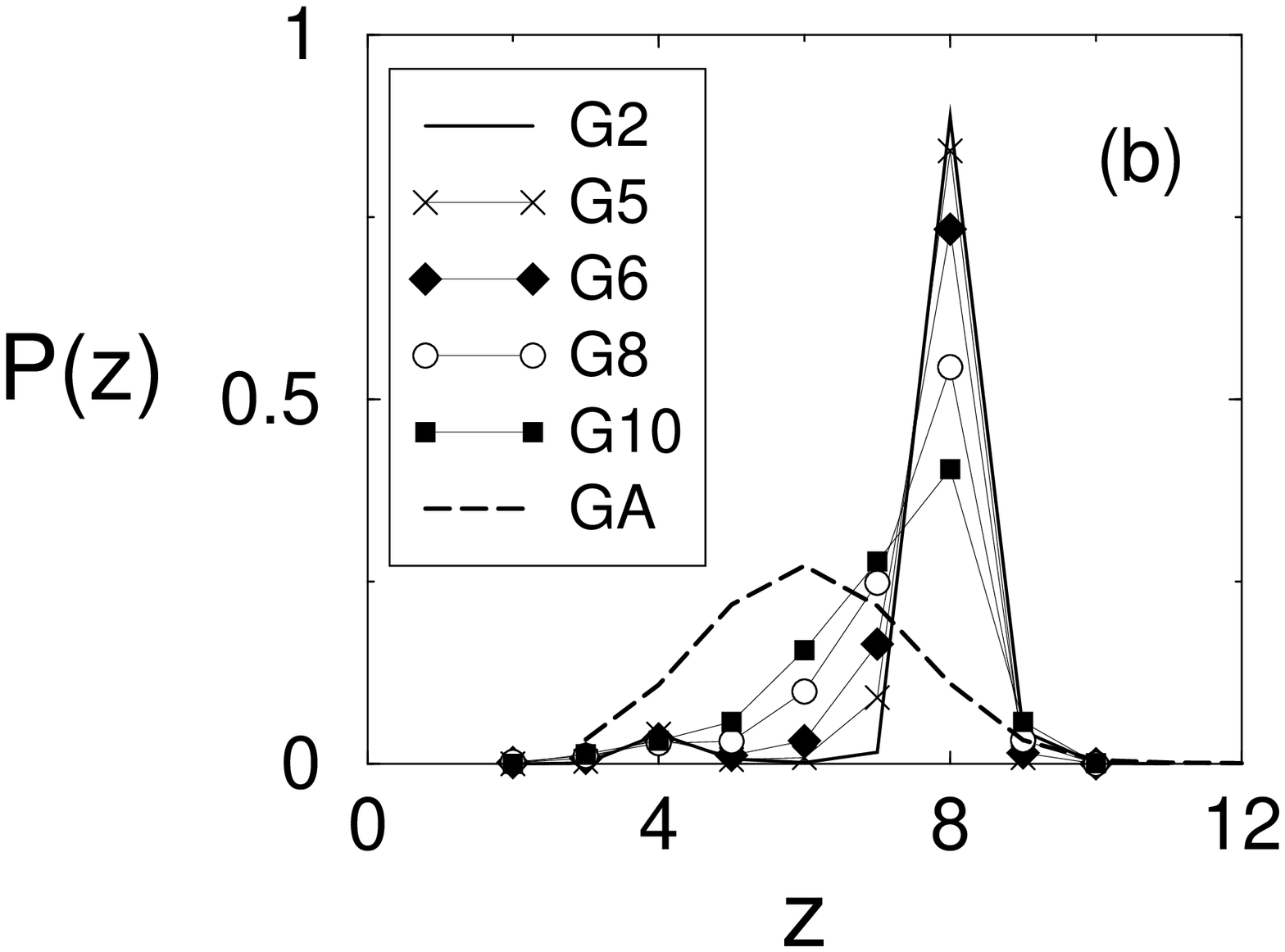}
\includegraphics[width=7cm]{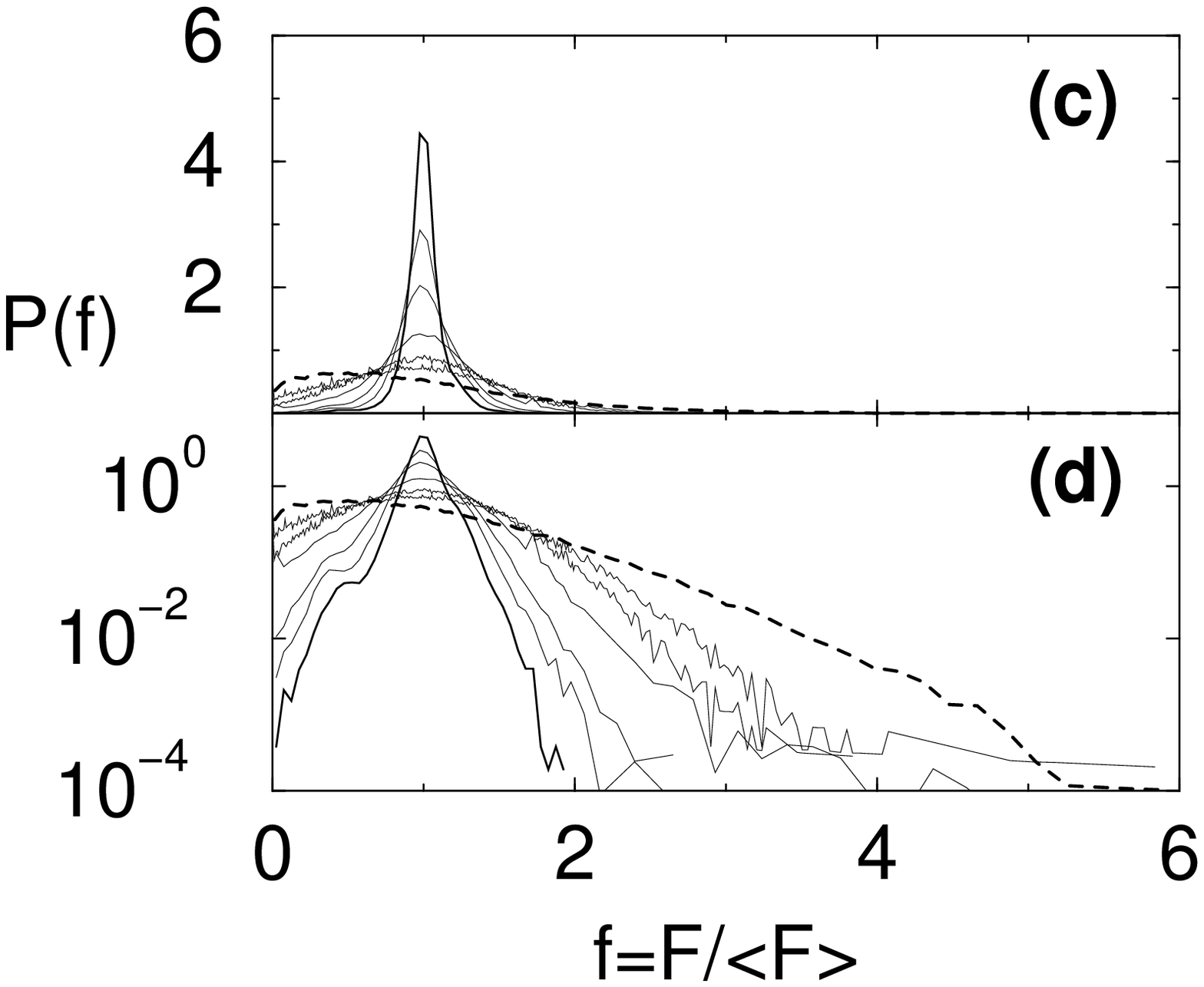}
\caption{Protocol (iii): Gravity sedimented packings. (a) Dependence
  of the depth-averaged packing fraction $\phi$, on $\delta N$. (b)
  Coordination number distributions $P(z)$, for different
  configurations, labelled with increasing disorder. Configuration
  $GA$ is an amorphous packing, with a packing fraction close to the
  random close packing value $\approx 0.64$. (c) and (d) are the
  distributions of normal contact forces $P(f)$, for varying disorder
  on linear axes and linear-log respectively. The thick dashed line is
  $GA$.}
\label{fig7}
\end{figure}

To provide a comparison with the quenched periodic packings already
presented, results for gravity-sedimented, frictionless spheres of
protocol (iii), are shown in Fig.~\ref{fig7}.
There are a number of subtle differences between the coordination
number and force distributions between protocols (i),(ii) and protocol
(iii). These differences primarily arise as a result of two effects.
Firstly, the gravity packings are not periodic in all three directions,
so surface effects play a role. Secondly, the gravity packings are
able to adjust their packing fractions with increasing disorder, as
shown in Fig.~\ref{fig7}(a), whereas the fully periodic system are at
a fixed packing fraction. However, the generic features of the $P(z)$
(Fig.~\ref{fig7}(b)) and $P(f)$ (Fig.~\ref{fig7}(c)) distributions are
similar. As more disorder is introduced into the lattice structure,
the distributions broaden. In the case of $P(f)$, the high-$f$ tail
becomes increasingly exponential. Thus, the fully periodic systems
capture the essential features of the more realistic gravity packings.
  
The recent focus on $P(f)$ has been emphasised due to suggestions that
particular properties of $P(f)$ can be used to signal the onset of
glassiness in a glass-forming system, and likewise, the approach of
the {\it jammed}, static state in a granular material or dispersion
\cite{ohern2}. The concept being that the development of a peak in
$P(f)$, for $f<1$, represents the balance of forces required for
mechanical stability into the jammed state - development of a yield
stress. For finite-temperature systems, the jamming transition of
purely repulsive, particles is accompanied by the development of a
peak in $P(f)$ at small $f$, as the temperature is lowered through the
glass transition temperature \cite{ohern1}.  (This picture is not so
clear for the case of systems with longer-range attractive forces, as
in Lennard-Jones systems \cite{glotzer1,sastry3}.) At
zero-temperature, the peak in $P(f)$ flattens into a plateau as the
density of a jammed packing is lowered towards the jamming transition
packing fraction \cite{makse1,leo17}.  Therefore, for
purely-repulsive, finite-range interactions, this jamming picture
seems to apply.

\begin{figure}[!]
\includegraphics[width=7cm]{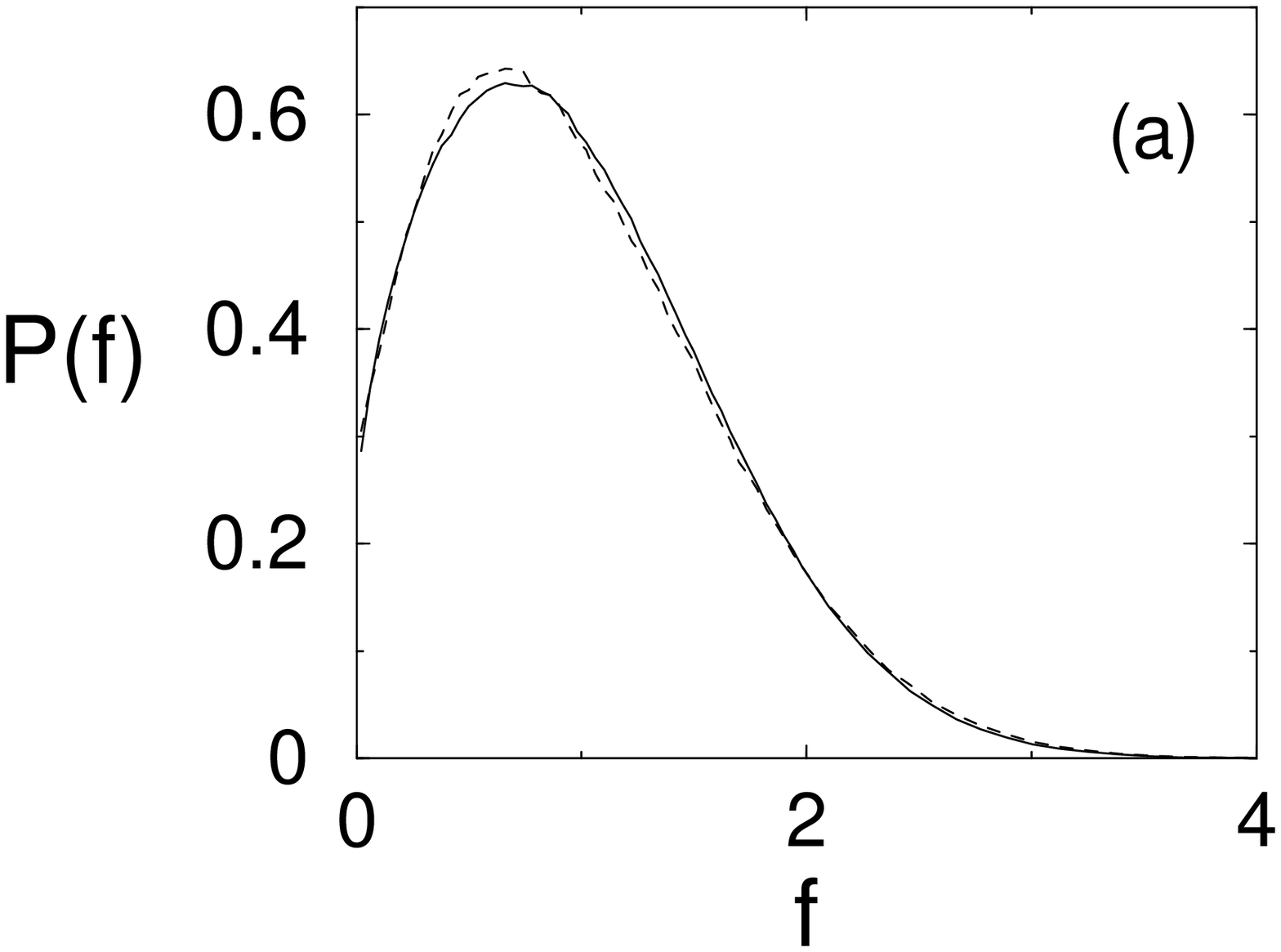}
\includegraphics[width=7cm]{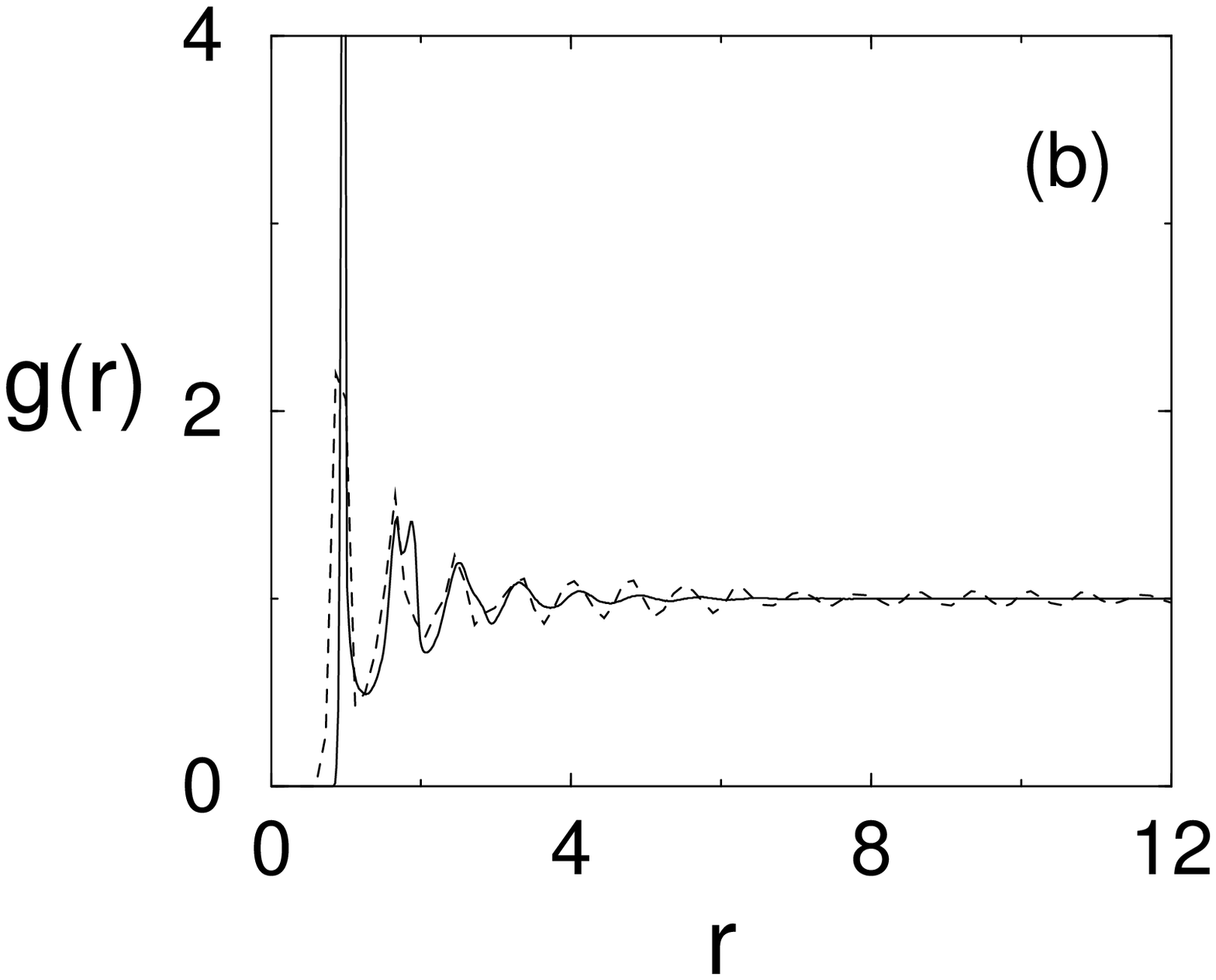}\\
\includegraphics[width=6cm]{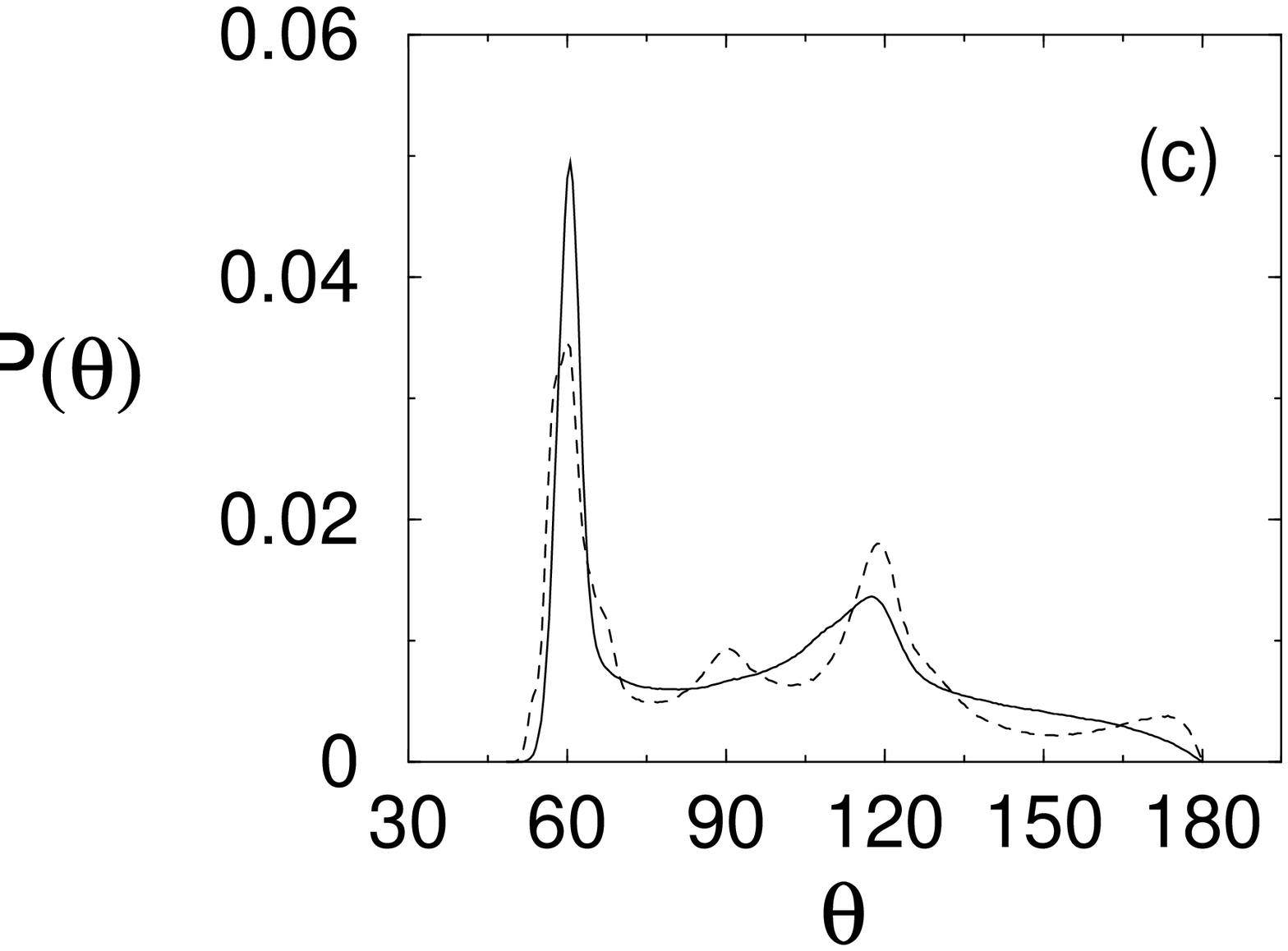}\\
\caption{(a) The distribution of contact forces $P(f)$ for
  purely-repulsive, soft-spheres at $T=0$ and $\phi=0.742$, with
  different histories. Soft-sphere glass (solid line) and partially
  melted crystal (dashed line). From $P(f)$ and visualisation of the
  force networks, both configurations appear very similar. Structural
  measures show, however, that the partially melted crystal is
  significantly more ordered than the glassy state: (b) The radial
  distribution function, $g(r)$, shows long-range oscillations and,
  (c) the three-particle contact angle distribution $P(\theta)$,
  contains additional structure indicative of ordering.}
\label{fig8}
\end{figure}

In Fig.~\ref{fig8}, results are shown for two jammed states of
purely-repulsive systems, each with $N=256000$ particles. One
configuration was generated using a fast molecular dynamics quench,
from a high (liquid) temperature to $T=0$. The other is a partially
melted crystal quenched back down to $T=0$. Despite the fact that not
only are both of these systems jammed and that the $P(f)$ curves in
Fig.~\ref{fig8}(a) sit on top of one another, differences in their
structures are evident from the radial distribution function $g(r)$
and $P(\theta)$ of Figs.~\ref{fig8}(b) and (c).This highlights the
fact that amorphous and quasi-ordered systems can exhibit similar
features in the jammed state.

\begin{figure}[!]
\includegraphics[width=7cm]{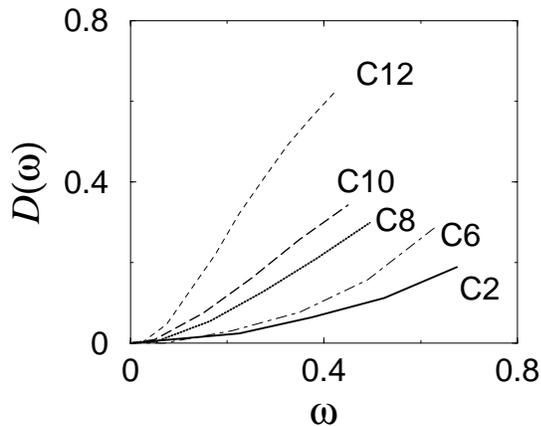}
\caption{Distribution of the vibrational normal modes of frequency
  $\omega$ - the density of states - $\mathcal{D}(\omega)$, for the
  quenched packings of protocol (i). Only, the low-$\omega$ region of
  the distribution is shown. With increasing disorder, the number of
  low-frequency modes increases relative to the fcc lattice.}
\label{fig9}
\end{figure}

Increasing disorder affects the mechanical properties of the packings
subject to external perturbations \cite{leo16}. Yet, for small amounts
of disorder, one expects the configurations to vary only slightly in
their properties from the underlying crystal. This is, indeed, the
case, as described by Eq.~\ref{eq2}. To determine how the disorder
influences the dynamical properties of the packings, the low-frequency
portion of the ``phonon'' density of states $\mathcal{D}(\omega)$,
were extracted for the $N=16384$ systems \cite{arpack1}. Figure
\ref{fig9} shows $\mathcal{D}(\omega)$ as a function of frequency
$\omega$, for configurations with varying amounts of disorder.
Increasing disorder leads to an increasing population of the
low-$\omega$ region in $\mathcal{D}(\omega)$, not unlike
lattice-disorder models \cite{schirmacher1,elliott2}.

\begin{figure}[!]
\includegraphics[width=7cm]{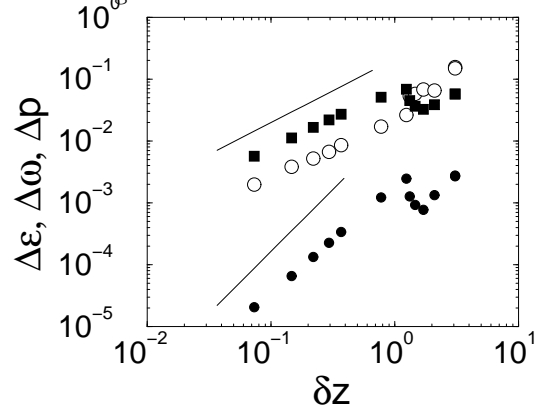}
\caption{Lowest normal mode frequency $\Delta\omega$ (open circles),
  pressure $\Delta p$ (solid squares), and contact energy $\Delta
  \epsilon$ (solid circles), relative to the original fcc values, as a
  function of the degree of disorder, measured by $\delta z$. The
  solid lines are power-laws with exponents one and two.}
\label{fig10}
\end{figure}

The relation, Eq.~\ref{eq2}, provides a useful measure of the
disorder, when the disorder is weak. Figure \ref{fig10} shows that
over this same range in $\delta z$, where the configurations are not
too different from the original lattice, many of the packings'
properties scale. In particular, {\it relative to the crystal state},
the lowest normal mode frequency varies approximately linearly with
coordination number. This is reminiscent of the way jammed amorphous
packings behave as the density is lowered towards the random close
packing point from above \cite{leo14,wyart1}. It is suggestive,
therefore, that there may exist a characteristic length scale
associated with the increasing disorder, though, as yet, one has not
been identified here.

In conclusion, the effect of structure on the properties of static
packings has been studied. Structure plays a dramatic role in
modifying the distribution of normal contact forces $P(f)$ of
frictionless particle assemblies, from an initial delta-function, for
the face-centred cubic array, to the more familiar `exponential'
decay, with finite disorder. Likewise, the distributions of the
coordination number broaden quickly. These two findings may, in part,
explain the reason why it has been difficult to {\it experimentally}
observe any significant dependence of $P(f)$ on structure
\cite{nagel5}. Even a relatively small number of defects can broaden
$P(f)$ quite substantially.

Dynamical properties of the packings have been investigated by
extracting statistics on the lowest-lying normal modes. As more
disorder is imposed, there is an increase in the density of states,
$\mathcal{D}(\omega)$, at small frequency, $\omega$. Provided the
structure remains close to the underlying lattice, the change in the
coordination number relative to the original lattice provides a useful
measure of the disorder. When the disorder is weak, vis-a-vis
Eq.~\ref{eq2}, the value of the lowest-lying normal mode frequency
scales approximately linearly with the disorder. It is somewhat
amusing that these relationships are not too different from what is
found in fully disordered packings near random close packing
\cite{wyart2}, where fractional changes in density play a similar role
as disorder does here. Between these extremes of densely packed
ordered arrays and loose amorphous packings, lies an intermediate
regime that is not so well characterised. This intermediate regime
retains a large degree of the contact topology of the original
lattice, yet exhibits a strong degree of heterogeneity in the contact
forces.

\acknowledgments 

I thank Moises Silbert for a critical reading of the manuscript.  Part
of this work was undertaken at the James Franck Institute, University
of Chicago, with grant support from Grant Nos.~NSF-DMR-0087349,
DE-FG02-03ER46087, NSF-DMR-0089081, and DE-FG02-03ER46088, and is
gratefully acknowledged.


\begin{thebibliography}{27}
\bibitem[{com({\natexlab{a}})}]{comment2}
\bibinfo{note}{The nature of the decay of the tail to $P(f)$ actually depends
  on how compressed the packing is \cite{makse1,leo17}.}
\bibitem[{com({\natexlab{b}})}]{comment3}
\bibinfo{note}{For the largest amount of disorder, there is a non-monotonic
  feature in the sense that the tail of $P(f)$ appears more Gaussian (decays
  more rapidly) at these high compressions. If the pressure, or packing
  fraction, of the amorphous packing was subsequently decreased, the high-$f$
  tail would again become increasingly exponential and is the reason for the
  differences between the circles and dot-dash line in Fig.~\ref{fig5}(b). See
  \cite{comment2}.}
\end{thebibliography}
\end{document}